# Understanding Idea Creation in Collaborative Discourse through Networks: The Joint Attention–Interaction–Creation (AIC) Framework

Xinran Zhu, Bodong Chen
xrzhu@upenn.edu, cbd@upenn.edu
University of Pennsylvania

**Abstract:** In Computer-Supported Collaborative Learning, ideas generated through collaborative discourse are informative indicators of students' learning and collaboration. Idea creation is a product of emergent and interactive socio-cognitive endeavors. Therefore, analyzing ideas requires capturing contextual information in addition to the ideas themselves. In this paper, we propose the Joint Attention–Interaction–Creation (AIC) framework, which captures important dynamics in collaborative discourse, from attention and interaction to creation. The framework was developed from the networked lens, informed by natural language processing techniques, and inspired by socio-semantic network analysis. A case study was included to exemplify the framework's application in classrooms and to illustrate its potential in broader contexts.

## Introduction

One of the central interests in Computer-Supported Collaborative Learning (CSCL) research is collaborative learning through discourse. Collaborative discourse, as a group practice, requires learners' maintenance of intersubjectivity via understanding and interacting with each other (Medina & Stahl, 2021). In digital environments, it is often presented as texts generated by learners. In CSCL, collaborative discourse is conceptualized in various ways. Scardamalia and Bereiter (2014), for instance, emphasized the role of learners' collective responsibility in knowledge-building (KB) discourse and regarded collaborative discourse as the process of collaborative knowledge creation in which each student assumes responsibility for contributing to the community's knowledge advancement. As a collaborative endeavor, learners participate in discourse by sharing ideas and continuously improving them through interaction (Scardamalia & Bereiter, 2014). Prior research has investigated socio-cognitive configurations for collaborative discourse in various contexts (e.g., Chen et al., 2022; Fischer et al., 2013). Results show that collaborative discourse promotes higher-order competencies such as argumentation (Noroozi et al., 2018) and problem-solving skills (Baker & Lund, 1997).

In CSCL, one goal of collaborative discourse is to create and improve ideas (Scardamalia & Bereiter, 2014). For example, in productive KB classrooms, ideas are generated around *authentic* problems; they are *improvable*, *diverse*, and can be *risen above* to become bigger ideas (Scardamalia & Bereiter, 2014). Researchers are taking different approaches to exploring characteristics of ideas gendered from collaborative discourse. Zhu and colleagues (2017) studied the distribution of ideas ("idea evenness") within KB communities. In another study, Chen et al. (2015) focused on the semantic feature of ideas by examining the "idea promisingness" in collaborative discourse.

One assumption of these efforts is that ideas generated by students are informative indicators of their learning and collaboration. Ideas are often stored as texts, such as discussion posts or chat messages. Many different approaches have been employed in the literature to analyze texts in collaborative discourse, including human-coded content analysis and computational methods (e.g., Park et al., 2021). Although investigating ideas can provide insight into learners' engagement by delving into their learning productions (e.g., discussion posts), it cannot provide a complete picture of where and how the idea originates. In CSCL, idea creation is not an isolated or momentary incident. Rather, it stems from emergent and interactive endeavors. Therefore, to understand idea production, we need comprehensive contextual information that captures idea formation, transactivity among ideas, and idea uptakes, using methods such as temporal analysis and sequential analysis (Ohsaki & Oshima, 2021; Suthers et al., 2010).

Networks offer a powerful tool to conceptualize, represent, and analyze collaborative discourse. In the context of CSCL, researchers have studied social interactions among learners or communities (De Laat et al., 2007). Similarly, networks can also be used to illustrate concepts that are linked semantically (Chen et al., 2021; Ohsaki & Oshima, 2021; Popping, 2000). One example is the modeling of collaborative discourse as socio-semantic networks. Socio-semantic network analysis (SSNA) captures not only the relational events among learners but also the emergent ideas inside the community's discourse by operating social relationships on top of a semantic layer (Poquet & Chen, in press; Ohsaki & Oshima, 2021).

These efforts in network sciences have shown promising directions to understand the development of ideas, but we may also want to know what happened before the ideas are created. Analyzing idea products alone may obscure the situated nature of collaborative discourse and may fail to adequately record the indexicality and context of learning products (Suthers et al., 2010). For example, the effectiveness of a collaborative dialogue cannot be assessed solely based on the end products (e.g., ideas from a forum post), we also care about whether the ideas originated from a clique of student groups or were shared by the whole class, and whether the ideas were centered around a specific topic or scattered across multiple ones. Those are all important questions to understand how ideas are created and improved in collaborative discourse.

This leads to the need of developing comprehensive analytic frameworks to capture the emergent and interactive nature of ideas in collaborative discourse by answering the question: *How are ideas generated in collaborative discourse?* Building on previous work on SSNA, we aim to answer this question by revealing contextual information about ideas creation in collaborative discourse, using networked methods and natural language processing (NLP) techniques. We first capture learners' *joint attention* before an idea is created through a *Joint Attention Network* (AN). Joint attention is a key feature of social cognition and an important construct in CSCL (Siposova & Carpenter, 2019; Wise et al., 2021). It helps groups to establish a common ground, take the perspective of peers, build on ideas, express empathy, or solve a problem together (Schneider et al., 2016). Studies have examined joint attention in various contexts, such as through eye-tracking techniques (Schneider et al., 2016). However, very little is known about joint attention in asynchronous collaborative discourse in text-based discussion environments. Second, we construct the *Interaction Network* (IN), which examines the direct interaction relations between learners. One hypothesis is that being in a center/influential position of IN leads to diverse ideas generation. Third, we construct the *Creation Network* (CN), which captures each individual student's position in the semantic network of ideas. This will offer insights into students' connectivity in the semantic spaces. Each of the three networks will provide distinct information about learners and their discourse.

## The AIC Framework for Collaborative Discourse

The AIC framework is grounded in CSCL literature and developed from the networked lens. In line with CSCL, learning is defined as the process of creating collectively valued knowledge by learners' contributing and improving ideas. From a network perspective, the analysis focuses on social ties formed through "interpersonal interactions surrounding the emergent network of ideas and knowledge structures" (Poquet & Chen, in press). The construction of networks considers multiple levels of connections, including the social, idea, and knowledge levels, as well as ties between and across them (Poquet & Chen, in press).

### Joint attention network

For collaborative discourse in digital spaces, joint attention refers to a shared focus on a common *reference*, for instance, a sentence from a shared reading. It supports the establishment of common ground and mutual knowledge, which is theorized as an important mechanism to help collaborators coordinate with one another (Wise et al., 2021). Although the concept of joint attention has been investigated through visual cues (e.g., eye gaze), it has been overlooked in the analysis of collaborative discourse in CSCL. To address the gap, AN investigates learners' attention by looking at the reference from the learning material (e.g., a sentence from a classroom reading), which intrigues the discourse. Informed by the network sciences and NLP, AN operates collaborators' attentions as a network of "semantic spaces" (Lund & Burgess, 1996) by examining the semantic similarity between each pair of references. Within the network, jointness is defined as having discourse around *common* or *similar* references. A common reference indicates that learners are having conversations about the exact same text (if the same text appears multiple times, we treat them as common references). A similar reference indicates that learners are having conversations about different but semantically similar texts. In summary, it answers the "where" question by revealing the origins of the ideas.

### Interaction network

With joint attention, learners start sharing ideas and building on each other's contributions through interaction, which is then captured by IN. IN tells us in the most intuitive way how learners are connected—students are connected if they have direct conversations during the learning activity, such as a response to a peer's writing reflection. In CSCL, direct interaction promises various levels of cognitive engagement via exchanging and negotiating ideas. Informed by Social Network Analysis (SNA), IN tells us important information about the individual student's contribution (e.g., by calculating degree centrality) as well as the class interaction patterns.

## Creation network

While IN provides a descriptive report of learners' interactions, CN then captures the production generated from the interaction. Building on the work in SSNA, one assumption of CN is that learners' ideas can be represented as clusters of keywords in the discourse (Ohsaki & Oshima, 2021). As such, CN represents learners' socio-cognitive engagement through a network of words. Words are selected from the discourse based on their influence. From all selected words, learners are connected if they use the same word. The hypothesis here is that students in the center position of the creation network tend to share similar ideas with peers, or "common knowledge." Students in marginalized positions tend to generate unique ideas.

## Situating AIC in collaborative discourse

Our AIC framework represents collaborative discourse as networks of three areas: joint attention, interaction, and creation. In a typical collaborative discourse in digital spaces, learners usually start by negotiating the meanings of shared textual learning material. In the case study presented in a later section where students participated in a social reading activity supported by web annotation technologies, students started by reading an assigned article on a specific topic in order to make sense of the learning object collaboratively. In this phase, learners interacted with the learning materials independently by highlighting some quotes and making annotations. Then, they read peers' annotations, replied to each other directly, or addressed others' annotations in a separate thread. AIC can capture the students' highlights (as the discourse reference) to construct AN, record interactions to construct IN, and create CN by extracting words from their annotations. As a whole, AIC captures the social reading and writing processes to understand the discourse through networks. The constructed networks can be further visualized and compared to provide insights into key concepts that are shared—or not—among groups (Popping, 2000).

## **Modeling networks within AIC framework**

Modeling collaborative discourse contains a series of analytical decisions in terms of retaining certain information and discarding the rest. In the context of collaborative discourse, we may ask: *What would be the unit of analysis, a word, a sentence, or a conversation unit? How do we define semantic similarity?* We propose an iterative heuristic approach to network construction explained below.

AIC leverages NLP techniques to analyze the texts as a corpus of semantic units. A wide range of NLP approaches has been used in CSCL under the assumption that language plays an important role in learning (McNamara et al., 2017). Dascalu et al. (2013) proposed a cohesion-based analysis model to access discourse by utilizing a series of NLP techniques, such as Latent Semantic Analysis (LSA) and Latent Dirichlet Allocation (LDA) for topic modeling. Later, Sullivan & Keith (2019) developed a theoretical framework for why and how an automated NLP approach can support microgenetic analysis of collaborative discourse.

To explain network modeling, let's recall the social reading example mentioned earlier. Each discourse activity, for example, one reading assignment, includes the *learning materials* (e.g., the reading) and the *learning artifacts* (e.g., the annotations). A *quote* from the learning materials is the reference of a certain learning artifact (e.g., an annotation made to a quote). It can be one or a couple of sentences. To construct AN, we adopted a pre-trained sentence-transformers model which produced 768-dimensional vectors for each quote. We then calculated pairwise cosine similarities for each pair of quotes. We defined joint attention as having discussions around a pair of quotes that had similarity scores equal to or higher than .8, as "joint quote". This threshold was chosen after experimenting with multiple configurations and checking the original text data. In the network, nodes are students. If two students have joint attention on at least one pair of quotes, they are connected through an edge. The network is undirected (for simplicity) and weighted. The weight is the sum of the similarity scores between each pair of joint quotes. Figure 1 illustrates the network construction mechanism. For example, Students A and B are connected by an edge with a weight of .8 because the quotes they highlighted (Quotes 1 & 2) have a similarity score of .8. Students B and C are not connected since their quotes are not semantically similar (.5 < .8).

**Figure 1**
*Example of AN Construction*

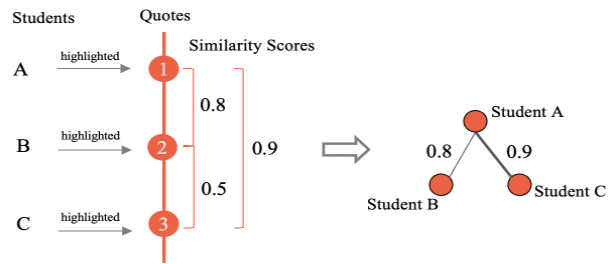

To construct IN, SNA was conducted to depict learners' interaction patterns. Similar to AN, we treat learners as nodes and direct conversations as edges. The network is weighted. The weight is the frequency of connections between two nodes. The network is undirected for simplicity.

To construct CN, we first utilized basic text mining techniques to tokenize and lemmatize the text from the discourse. Next, part-of-speech taggers were applied to extract only nouns that are most likely to represent the substantive content (Bail, 2016). For each dialogue (e.g., on a reading), we extracted words from conversations that occurred at least five times. We then measured the term frequency-inverse document frequency (tf-idf) to identify word significance within the corpus. Tf-idf is used to weigh a keyword in the corpus and assign importance to that word based on its frequency (Aizawa, 2003). Different from term frequency, tf-idf considers words that are important but rare in a given corpus. Based on tf-idf, we removed the five words with the lowest scores, such as "dance" in a class about dance history, or "learning" in a learning sciences class. We then extracted the top 70 words. There might be duplicates here since the same words in different annotations can have different tf-idf scores. Duplicates were removed only if they belong to the same author for simplicity. These analytical decisions were made based on the concept of learning constructs and the context of the discourse data. SSNA was then applied to get a two-mode socio-semantic network (see Figure 2a). We then projected the network to the learner-learner level (see Figure 2b). Students are connected if they use at least one same word. The network is undirected for simplicity. The weight of edges is the frequency of co-occurred words between nodes.

**Figure 2**
*Example of CN Construction*

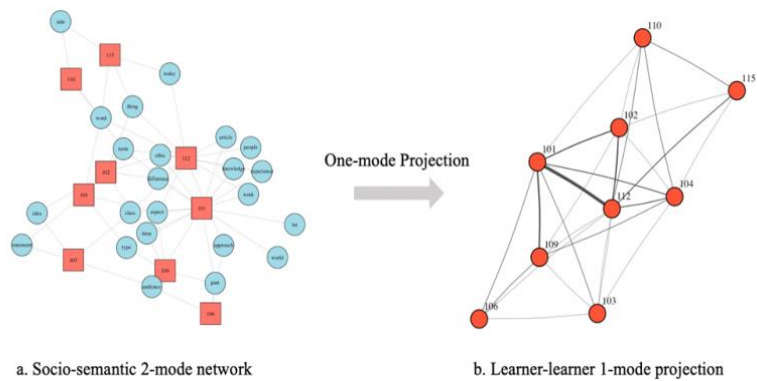

a. Socio-semantic 2-mode network       b. Learner-learner 1-mode projection

## Case study: Applying AIC to an analysis of collaborative discourse

### Context and data

To exemplify the AIC framework, we examine collaborative discourse in an undergraduate Dance History class offered by a large public university in 2020. This case study drew on a secondary dataset from a design-based research project that aims to support social reading through web annotation technologies. In this course, students ($n = 13$) were engaged in reading course materials, participating in weekly virtual meetings, and writing reflective essays. For the reading activities, the instructor used a web annotation tool named *Hypothes.is* to support collaborative sensemaking via social annotation. Throughout the semester, students were required to read 1-2 readings each week, annotate each reading, and reply to each other's annotations. By annotating the readings and replying to peers' annotations, students created ideas and built on each other's ideas. Over one semester, the class generated 478 Hypothes.is annotations and 469 replies in 18 readings across 11 weeks. On average, each reading had 26.6 annotations (SD = 2.6) and 26.1 replies (SD =4.0). Table 1 provides descriptive statistics of student annotations, replies, and their average words per annotation. To illustrate the framework, the AN, IN, and CN networks were constructed for two selected readings, 3a, and 6b. These two readings were selected based on our previous analysis of students' knowledge construction levels. Knowledge construction levels for students in Reading 3a were among the highest, indicating a higher cognitive engagement, while students in Reading 6b demonstrated a lower cognitive engagement (Zhu et al., 2021).

**Table 1**
*Descriptive Statistics of Discourse Data*

| Reading | 1 | 2a | 2b | 3a | 3b | 4 | 5 | 6a | 6b | 7a | 7b | 8 | 9a | 9b | 10a | 10b | 11a | 11b |
|---|---|---|---|---|---|---|---|---|---|---|---|---|---|---|---|---|---|---|
| Posts | 28 | 27 | 32 | 27 | 26 | 29 | 28 | 28 | 28 | 24 | 26 | 29 | 29 | 23 | 22 | 24 | 26 | 22 |
| Replies | 25 | 28 | 24 | 27 | 24 | 38 | 27 | 30 | 23 | 28 | 27 | 27 | 26 | 25 | 22 | 21 | 28 | 19 |
| Average words per post | 23.9 | 26.2 | 22.7 | 24.5 | 23.4 | 24.4 | 24.5 | 26.1 | 23.1 | 27.1 | 24.8 | 25.1 | 22.4 | 21.3 | 25.2 | 21.7 | 25.2 | 24.8 |

## Analysis

AN, IN, and CN networks were constructed for two selected readings, 3a, and 6b. The constructed networks provide intuitive visualizations of how learners were having joint attention, interacting, and contributing to the community's knowledge. To further analyze individual students' participation and engagement at the class level, we then calculated network measures at the network level and node level.

At the network level, we wanted to know: *Are students' attentions concentrated or distributed more evenly across the reading? Is the class discourse dominated by a few students? Are students' ideas focused or scattered in various directions?* To answer these questions, we first computed the clustering coefficient (or transitivity) for AN. In networks, the clustering coefficient is calculated based on triads and it is a measure of nodes' tendency to group together with dense connectivity (Carolan, 2013). It can provide information about the distribution of joint attention in discourse. A high clustering coefficient implies a high level of joint attention on one or a few quotes, which might help establish the common ground and take the perspective of peers in collaborative discourse (Schneider, et al., 2016). Similarly, clustering coefficients were also applied to CN to understand the distribution of words. Finally, we measured the centralization of IN based on the degree centrality of nodes to understand if the discourse was dominated by a few learners or distributed evenly.

At the node level, we wanted to know: *Is a student having extensive joint attention with peers or focusing on unique parts of the reading? Is a student actively interacting with all peers through replies? Is a student more likely to use the same words with peers in the annotations?* To answer these questions, we first measured the *closeness centrality* for each node in AN. Closeness centrality measures how close a node is to all other nodes in the network. Being "close" to others in AN implies having a higher level of joint attention through direct and indirect interactions. We then calculated the *betweenness centrality* for each node in IN and CN. Similar to closeness centrality, betweenness centrality also reflects students' positions and influence in the network. Instead of measuring the distance, betweenness centrality captures the number of times a node connects to other nodes. In IN, higher betweenness centrality indicates a powerful position in the network by bridging conversations and mediating interactions. In CN, students with high betweenness centrality tend to produce similar ideas with various peers. Both closeness and betweenness measures were normalized to compare across networks. These centrality measures are used in CSCL widely to understand collaboration and interaction.

## Findings

Figure 3 presents the resulting networks of students for Reading 3a (in red) and 6b (in green). Isolated nodes were excluded. As described earlier, In AN (Figure 3a), students are connected if they have joint attention, such as making annotations or replies to the same quote or annotating different quotes with highly similar semantic meanings. If the edge is thick (e.g., the edge between 102 and 112 in Reading 3a), the students share attention to multiple quotes. In IN (Figure 3b), students are connected if they have direct interactions. The weight of edges indicates how often they interact. In CN (Figure 3c), students are connected if they post the same keyword. Thicker edges imply that two students were using multiple same words in their annotations.

**Figure 3**
*The AIC Networks for Reading 3a & 6b*

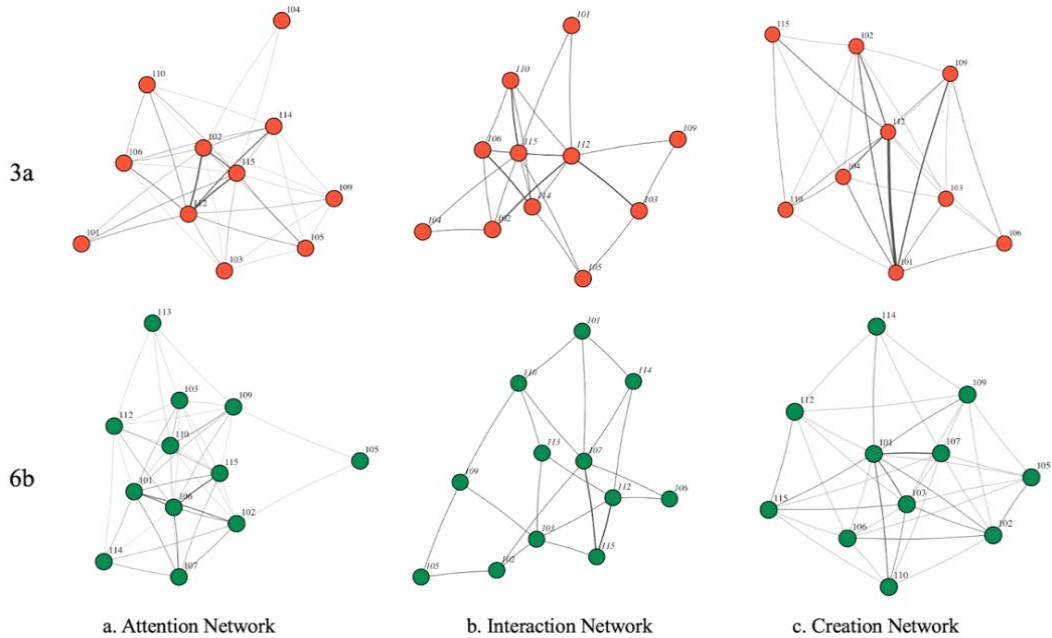

a. Attention Network    b. Interaction Network    c. Creation Network

The network presentations can first provide some intuitional observations about each reading activity. In Reading 3a, there were consistencies between AN and IN in terms of nodes' positions, which is reasonable since having interaction indicates joint attention. For example, Students 104 and 101 were in relatively marginalized positions in both networks. However, in Reading 6b, student 106 had joint attention with multiple peers (e.g., 101, 115, 107, and 102), but was relatively marginalized in the interaction network. This implies that student 106 tended to have joint attention with peers without direct interaction. From the creation network, student 106 used the same words with multiple peers. This can imply that student 106 reached common knowledge with peers without many direct interactions. Furthermore, looking at the ties, it is worth noting that students' relationships can differ across three networks. For example, in Reading 3a, students 112 and 103 had strong ties in IN, but relatively weak ties in AN and CN. It indicates that even though they had multiple direct interactions, they tended to have different foci and diverse ideas.

To demonstrate AIC's capacity to understand individual students' engagement, we present the node-level measures for reading 3a as an example in Table 2. Isolated nodes are presented as "na". Table 2 suggests that students were having different participation patterns across three networks. Student 112, for example, held influential positions in all networks, indicating high levels of joint attention and active interactions, resulting in sharing similar ideas with peers. Student 115, on the other hand, had the highest levels of joint attention and interaction but the lowest levels of betweenness centrality in CN, indicating that the student had common interests and active interactions with peers, while producing unique ideas.

**Table 2**
*Node-level Measures for Reading 3a*

| Student | 101 | 102 | 103 | 104 | 105 | 106 | 107 | 109 | 110 | 112 | 113 | 114 | 115 |
|---|---|---|---|---|---|---|---|---|---|---|---|---|---|
| AN Closeness | .56 | .63 | .63 | .56 | .67 | .67 | na | .67 | .63 | .91 | na | .71 | 1.00 |
| IN Betweenness | .00 | .02 | .02 | .00 | .11 | .05 | na | .03 | .01 | .26 | na | .05 | .36 |
| CN Betweenness | .05 | .05 | .02 | .05 | na | .00 | na | .02 | .01 | .12 | na | na | .00 |

The AIC framework can also provide network-level information. Table 3 shows the network-level measures for Reading 3a and 3b. For both readings, students had a high level of joint attention. Their discourse was relatively evenly distributed instead of centralized in a few clusters. Also, students have reached common knowledge. With the network-level information, we can also compare discourse based on their network structures. For example, as shown in Table 3, Reading 3a's CN transitivity score was higher than 6b, which is consistent with our previous study on their cognitive engagement level (Zhu et al., 2021). However, the IN centralization score for 3a was also high, which implies that the high engagement might relate to a few students' dominations in the discourse.

**Table 3**
*Network-level Measures*

| Reading | AN transitivity | IN centralization | CN transitivity |
|---------|-----------------|-------------------|-----------------|
| 3a      | .72             | .38               | .80             |
| 6b      | .74             | .23               | .72             |

## Discussions and implications

In this paper, we developed an analytic framework to capture the emergent and interactive nature of collaborative discourse by answering the question: *How are ideas generated in collaborative discourse?* AIC depicts students' engagement in collaborative discourse through three dimensions: joint attention, interaction, and creation. Grounded in the CSCL literature and inspired by the work of SSNA, the AIC framework captures the contextual information of collaborative discourse ranging from attention and interaction to creation.

To illustrate the potential of applying the framework to a broader context, we applied it to an example discourse dataset. As the findings show, AIC can be used to investigate network-level structures as well as individual node connectivity. The AN provides insights into where students' discourse comes from. In particular, it captures the distribution of attention as well as individual students' joint attention levels. The IN then tells us how students were interacting facilitated by joint attention. Similarly, it shows how the class is connected at the network level as well as individual student participation. Lastly, the CN captures the production out of attention and interaction. It sheds light on the structure of the class's productions – if ideas are concentrated or scattered. Besides, it provides insights into individual students' contributions to the community's knowledge advancement.

Collaborative learning is a complex endeavor. Therefore, integrating multiple networks to capture different dimensions is necessary. In AIC, multiple networks can be used together to inform class design or evaluation. For example, the instructor can direct students to interact with different peers if IN is sparse but CN is concentrated, to develop diverse perspectives. The AIC framework can also explain individual students' participation by looking at their positions in networks. For example, a student who is influential in IN can either share the same ideas or bring diverse perspectives. Similarly, a student who has a high level of joint attention with peers can either develop diverse perspectives or deepen thinking in one direction. In addition, the AIC framework can also provide insights for detecting emerging roles in the discourse. In CSCL research, roles have been recognized as a fundamental aspect of group dynamics, which is essential for collaborative knowledge construction (Ouyang & Chang, 2019). Emerging role detection can help adjust and calibrate the activity design in real-time to provide dynamic models of teaching and learning (Wise et al., 2021).

In conclusion, the proposed framework aspires to capture important dynamics in collaborative discourse using joint attention–interaction–creation networks. This paper makes the initial step to establish a proof-of-concept. Future work will further refine the NLP procedures by training the model in our context, applying the framework to other discourse contexts, combining AIC network analysis with other analytical methods, and integrating pedagogical designs (such as assigning roles in the discourse) in the baseline model.